\documentclass{appolb}
\usepackage{epsfig}
\usepackage{picinpar}
\newcommand{\fd}{f_D}
\newcommand{\fds}{f_{D_s}}
\newcommand{\piz}{\pi^0}
\newcommand{\gev}{\,\mathrm{GeV}}
\begin{document}
\title{Charm
\thanks{Presented at The XXVII Physics In Collision, June 26-29, 2007, Annecy, France}%
}
\author{Hanna Mahlke
\address{Laboratory of Elementary-Particle Physics \\
Cornell University, Ithaca, NY 14853, USA}
}
\maketitle
\begin{abstract}
New results on properties and decays of open charm and charmonium states
are reviewed. The emphasis is on examples that illustrate the various
aspects through which studies of charm physics impact the field.
\end{abstract}
%  12. 	Specific theories and interaction models; particle systematics
%12.38.Qk 	Experimental tests
%  13. 	Specific reactions and phenomenology
%13.20.-v 	Leptonic, semileptonic, and radiative decays of mesons
%13.25.Ft 	Decays of charmed mesons
%13.25.Gv 	Decays of J/psi, Upsilon, and other quarkonia
%13.66.Bc 	Hadron production in e-e+ interactions
%  14. 	Properties of specific particles
%14.40.-n 	Mesons

\section{Introduction}
The charm sector provides a unique testing ground for
understanding the strong interaction. The large data samples of open
charm and charmonium decays now available facilitate detailed
theory-experiment comparisons. Moreover, many lessons learned
from charm can be transferred to the bottom sector.
Perturbative methods are applicable when charm decay is concerned; 
yet it is also possible to explore relativistic effects due to
the light (in comparison with bottom) charm-quark mass.  

The immediate goals of the analyses presented here~\footnote{
In this document, reproduction of graphics shown in the talk 
will be limited to figures that are not readily available 
in publications.
}
are to study charm for its own sake, 
to treat it as a calibration ground for methods developed for heavier systems, 
and to use it as a production site for light-quark systems.

\section{Open Charm}

\subsection{Leptonic and semileptonic decays}
Leptonic and semileptonic decays allow the study
of the effect that QCD has in weak decays, where
the interaction between the quarks modifies the
decay rates induced by the weak process. 

For leptonic decays such as $D \to \ell \nu$, the
goal is a precision branching fraction measurement which,
via the relation $\Gamma(D^+ \to \ell^+ \nu )\propto
\fd^2 |V_{cd}|^2$,
can be turned into a measurement of the decay 
constant $\fd$, with the
external input of the CKM matrix element $V_{cd}$. 
Similar relations hold for leptonic $D_s$, $B$, and $B_s$ decays. 
Measurements of $\fd$ and $\fds$ can be compared to 
calculations, thereby validating computation techniques that can
then confidently be applied in the $B$~system, where entities
quantifying the effects of strong interaction
must be provided as an external input in order
to arrive at {\sl e.g.}~$V_{td}$. 
The experimental  precision currently achieved 
is about $5\%$ on $\fd$ (CLEO~\cite{cleo:fd}) and 
$8\%$ on $\fds$ (CLEO~\cite{cleo:fds} and BaBar~\cite{babar:fds}).
The CLEO measurements are absolute determinations, while the
BaBar result is obtained relative to the decay $D_s \to \phi \pi$,
which entails a dependency on the normalization branching fraction. 
A recent lattice QCD (LQCD) calculation~\cite{lqcd:fd_fds}
is in broad agreement with the measured results given the
experimental uncertainties,
but quotes a superior precision of $1-2\%$.

For semileptonic decays, for example $D \to \pi \ell \nu$,
a different kinematic variable enters, namely the momentum 
transfer $q^2$ to the outgoing hadron. The differential rate 
for decay to a pseudoscalar, $d\Gamma/dq^2
(D \to P \ell \nu)$, is proportional to $|V_{cq}|^2 \times
|f_+(q^2)|^2 \times p_P^3$. 
The modification of the weak decay by the strong force present
between the ingoing quark, outgoing quark, and spectator quark
is described by the pseudoscalar form factor $f_+(q^2)$.
The value of $p_P$ is determined by kinematics, and 
also the form factor shape does not require additional external 
information. For normalization, experiment can access the 
product $V_{cq} \times f_+(0)$, and therefore
input of either $V_{cq}$ or the form factor normalization $f_+(0)$ 
is required to determine the other one. 

Experimental progress has seen drastic improvements 
in accuracy for all branching fractions,
and some new ones observed for the first time, such as
$D^0 \to (K^- \pi^+\pi^-) e^+ \nu$~\cite{cleo:d_kpipienu} at $4\sigma$
statistical significance, 
${\cal B} (D^0 \to (K^- \pi^+\pi^-) e^+ \nu) = 
(2.8 ^{+1.4}_{-1.1} \pm 0.3) \times 10 ^{-4}$. 
The decay is seen to proceed dominantly through a $K_1^-(1270)$.

\begin{figure}[h]
\includegraphics*[width=\textwidth]{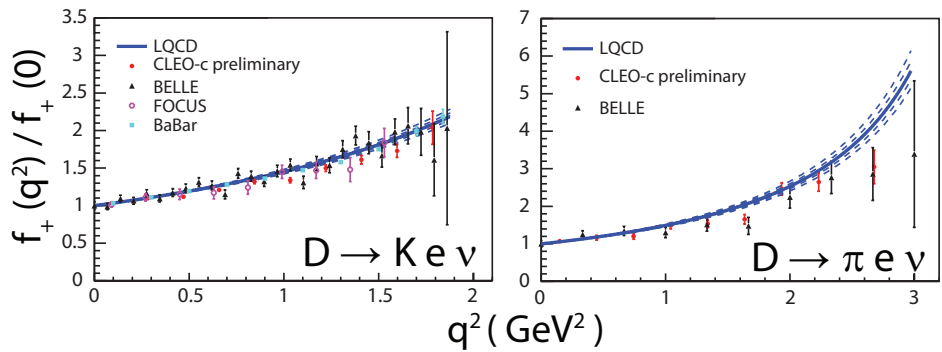}
\caption{World data on $D \to K \ell \nu$ and $D \to \pi \ell \nu$
form factors, with an LQCD prediction \cite{lqcd:ff_shape} overlaid.}
\label{fig:formfactor}
\end{figure}

A comparison between calculated and measured form factor normalizations
shows experiment still ahead of theory in terms of precision.
As to $D \to P$ form factor shapes, data currently can accommodate
the modified pole model~\cite{ff:mod_pole} or 
the series parametrization~\cite{ff:series}. World data
agree on the form factor shape within the current error levels;
the shape predicted from an unquenched LQCD calculation in 
Ref.~\cite{lqcd:ff_shape} agrees with the data for $D \to K \ell \nu$
(Fig.~\ref{fig:formfactor} left),
but seems to trend towards values larger than the data at upper $q^2$
in $D \to \pi \ell \nu$ (Fig.~\ref{fig:formfactor} right).

\subsection{Hadronic decays}
The hadronic decay rate can give insight into a more intricate 
interplay between manifestations of the strong interaction. 
An added complication is that, with several hadrons produced, 
final state interactions take place which are hard to quantify.

\begin{figwindow}[2,l,%
  \includegraphics*[width=2.5in]{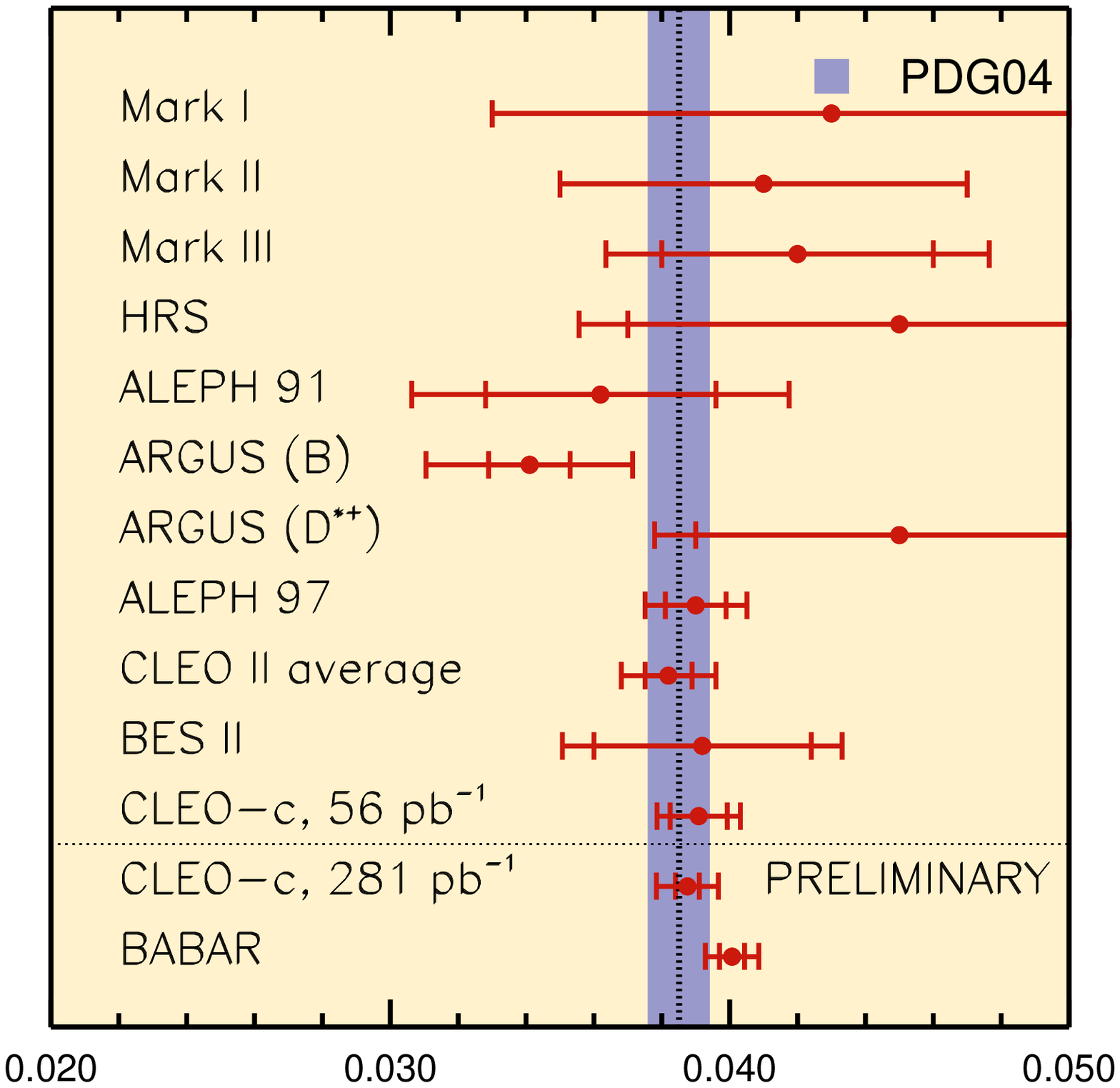},%
  {\label{fig:D_Kpi} Measurements of the branching fraction 
${\cal B}(D^0 \to K^- \pi^+)$.}] 
Branching fraction measurements have an especially far-reaching 
impact for channels that are used as normalization modes.
One such example is $D^0 \to K^- \pi^+$,
for which both CLEO~\cite{cleo:d0_kpi} and BaBar~\cite{babar:d0_kpi} 
have new measurements using very different techniques. 
The relative precisions achieved by both
are about 2\%, dominated by systematic uncertainties.
At this level of sensitivity, estimating the uncertainty
on final state radiation corrections is an important task
as these contributions are not negligible.
CLEO also measured absolute branching fractions for other Cabibbo-favored 
$D^0$ and $D^+$ decays~\cite{cleo:d0_kpi}: most of these results
are the most precise to date.
\end{figwindow}

\vspace{8mm}

A comparison of the rates for the
Cabibbo-favored decay $D^0 \to K^- \pi^+$ with
the Cabibbo-suppressed ones $D^0 \to \pi^- \pi^+$ and $D^0 \to K^- K^+$
found that, after adjusting for phase space, the measured 
rates~\cite{pdg06} do not behave as expected, namely
proportionally to the squares of the CKM matrix element ratios involved:
$( V_{cd} / V_{cs} )^2 \approx 0.05$ for 
$\pi^- \pi^+: K^- \pi^+$, $K^- K^+ : K^- \pi^+$,
and unity for $\pi^- \pi^+: K^- K^+$.
Instead, one finds, for
$ \pi^- \pi^+ : K^- \pi^+$, $K^- K^+ : K^- \pi^+$, and 
$K^- K^+ : \pi^- \pi^+$ the values~\cite{babar:threebody}
$0.034 \pm 0.001$, $0.111 \pm 0.002$, $3.53 \pm 0.12$.
Both Belle~\cite{belle:threebody} and BaBar~\cite{babar:threebody} extended 
this study to see if the same imbalance
would hold for three-body decays, in which a $\piz$ was added to
each of the above channels. 
Those measurement support the expectation at the level of $~30\%$.

CLEO~\cite{cleo:Ds_PP} 
performed a systematic search for ten 
$D_s \to P_1 P_2$ decays, where modes with 
           $P_1 = K^+$ or $\pi^+$ and 
           $P_2 = \eta,\ \eta, \ \piz, \ K_S$ or $K_L$ are studied.
This comprises all decays to a pair of mesons from the lowest-lying
pseudoscalar meson nonet.
The analysis measures Cabibbo-suppressed branching fractions relative 
to those of the corresponding Cabibbo-favored mode. The suppression ratio 
is, again, expected to be of order $|V_{cd} / V_{cs}|^2$, which is confirmed
by the data. Signals are seen in all modes except for the isospin-forbidden 
decay $D_s \to \pi^+\pi^0$. 

Other questions to be addressed with hadronic multi-body decays 
are the production of intermediate resonances and their properties, 
or suppression patterns of certain channels.
The description of the multi-body final state is dependent on the
formalism used and introduces model dependencies that cannot be removed.
Points of debate are the existence of a sound
theoretical basis (as opposed to just achieving
a satisfactory description of the data), quality control (what
constitutes a satisfactory description), knowledge of inferred or
expected intermediate states (in particular, consistency with
scattering experiments), and the importance and treatment of
final state interactions. Similar questions arise in the $B$~system.

FOCUS~\cite{focus:d_4pi} analyzed 
$D^0 \to \pi^-\pi^+\pi^-\pi^+$ events, first determining the
branching fraction relative to $D^0 \to K^-\pi^+\pi^-\pi^+$,
which is the most precise to date and in agreement with data 
from CLEO-c, but then moving to an amplitude analysis. 
Two important motivations, aside from the fact that
for this decay it had not been done before, are 
to examine the importance of final state interactions (which are 
is expected to be greater in four-body than in three-body decays);
and developing an understanding of intermediate
resonances such as the $a_1(1260)$, expected to be relevant in
$B^0 \to \pi^-\pi^+\pi^-\pi^+$.
The FOCUS model incorporates ten baseline components: 
$D^0 \to a_1(1260)^+ \pi^-$ with
$a_1(1260)^+ \to \rho^0 \pi^+$ ($S$ and $D$ wave) and
$a_1(1260)^+ \to \sigma \pi^-$,
$D^0 \to \rho^0 \rho^0$ in three helicity states,
and 
$D^0 \to \pi^+ \pi^- + {\cal R}$ with 
${\cal R}=\sigma, \ \rho^0, \ f_0(980), \ f_2(1270)$.
The fit returns the $a_1(1260)$ a dominant contributor with
a fit fraction of about $60\%$,
as is the case also in $K^-\pi^+\pi^-\pi^+$ and $K^0\pi^+\pi^-\pi^+$.
The group $\rho^0 \rho^0$ yields a fit fraction of 25\% 
the rest,
$\pi^+\pi^- + {\cal R}$, contributes about 11\%. While these
components get the gross features of kinematic distributions
right, the confidence level is low and cannot be improved by
inclusion of further signal amplitudes. This can either imply
that the model is too simplistic, and/or that final state interactions
cannot be ignored. 

The information on the mass and the width of the $a_1(1260)$ 
extracted from the fit result is
more precise than the current PDG average and agrees with theoretical
predictions (of similar precision).

\subsection{Open-charm spectroscopy}

The production of pairs of open-charm mesons can be observed in
$e^+e^-$ scattering, and particularly interesting features are
observed near threshold. 
The production cross-sections as a function
of center-of-mass energy show structure that can either directly
be interpreted as bound states in the charmonium system, 
or as the effect of interference from several resonance 
at nearby energies. Since the bound states decay to different
pairs of mesons depending on their quantum numbers and mass,
in order to obtain a complete picture it is important to measure 
the production cross-section for different pairs of mesons, not 
just the inclusive or just, for example, $D^+ D^-$. 
Predictions for the production of $D$~pairs can be found 
for example in~\cite{eichten:opencharm_xsect}.

Production cross-section data on exist from 
CLEO ($e^+e^- \to D \bar D$, $D^* \bar D$, $D^* \bar D^*$, 
$D_s^+ D_s^-$, $D_s^{*+} D_s^-$, $D_s^{*+} D_s^{*-}$ at
a center-of-mass energy $E_{CM} \sim 4\gev$) 
as well as Belle and BaBar ($e^+e^- \to \gamma + D \bar D$, 
$D^{*+} D^-$, $D^{*+} D^{*-}$ at $E_{CM} \sim 10\gev$; 
\cite{belle:opencharm_xsect}); both agree on the rough features.
A comparison between the sum of the CLEO channels with
an inclusive measurement (with the $uds$ continuum subtracted)
reveals multi-body contributions of the type $D^* D \pi$.

The characteristic enhancements in the inclusive cross-section 
between 3.7 and 4.5~GeV are commonly associated with the 
$\psi(3770)$, $\psi(4040)$, \linebreak
$\psi(4160)$, and $\psi(4415)$ states 
of charmonium. Their mass and widths were determined in earlier
experiments. A re-evaluation by BES of scan data in the region
of $2-5\gev$~\cite{bes:2-5gevscan} 
applying a model that takes interference between the resonances 
into account leads to, in some cases, substantially altered
parameters for the mass, width, and electronic coupling, owing 
to the fact that the three higher resonances are broad,
and, consequently, interference effects become visible.
The $\psi(3770)$ results are in line with earlier results
of more focussed scans (BES~\cite{bes:psi3770scan}). 

There are many other members of the charmonium system above
open-flavor threshold that
have not yet been observed. Many of them cannot be identified
in direct $e^+ e^-$ production because of their quantum
numbers; $p \bar p$ scattering or transition from a higher-mass
state offer additional avenues. It is important to continue
searching for those states in order to compare with and refine
predictions for the spectrum of quarkonium states. 

\section{Charmonium}

\subsection{Spectroscopy}

In contrast to the somewhat hazy
situation of charmonium states above open-flavor threshold,
all states below have been observed~\cite{pdg06}. 
Large samples exist for the $J/\psi$ and $\psi(2S)$,
which are well-studied. The experimental focus is now 
on comparing the two states, 
identifying rare decays, 
and investigating the resonant substructure in multibody states. 
As for open charm,
this provides information on the intermediate states produced
and gives insight into the decay dynamics.
The masses and widths are known well. A scan of the $\psi(2S)$
by E385~\cite{e385:psi2Swidth} 
led to the current best 
results on $\Gamma(\psi(2S))$ and $\Gamma(\psi(2S)) \times
{\cal B} (\psi(2S) \to p \bar p)$. 
The process used was $p \bar p \to \psi(2S)$ with 
$\psi(2S) \to e^+e^-$ or $\psi(2S) \to X J/\psi \to X e^+e^-$. 
The analysis makes use of the beam
energy spreach that is comparable to the structure investigated
as opposed to the ~MeV range that $e^+e^-$ machines are limited to. 

The $\chi_{cJ}$ states can be studied using the reaction
$\psi(2S) \to \gamma \chi_{cJ}$, where they are produced
at a branching ratio of order $10\%$ each. 
This implies that the $\chi_{cJ}$ 
data are not far behind the $\psi(2S)$
in statistical power, and similar studies as for the $\psi(2S)$
are being conducted.
Once the transition photon is identified, the $\chi_{cJ}$
are easy to handle experimentally. 
The transition rates are affected by relativistic corrections,
and thus measuring them accurately is important to guide theory.
A variety of approaches and models exist; the experimental
precision is currently well below the spread of 
predictions~\cite{e1_trans}.
The $\eta_c(1,2S)$ and the $h_c$ are less well known, and studies
to learn more about their properties and decays are underway.

\subsection{Hadronic decays}
Based on the observations that to date much fewer radiative decays have
been observed for the $\psi(2S)$ than for the $J/\psi$, and that the
ones seen all have branching fractions at the level of $10^{-4}$
or $10^{-5}$ while a naive scaling leads to the expectation of about
$1\%$ for the sum $\psi(2S) \to \gamma gg$, BES executed a survey
of multi-body decay modes with pions and kaons~\cite{bes:radpsi2S}. 
They found many
more signals, but all in the same range as the ones previously measured.
This raises the question whether modes with even higher multiplicity
matter as much as to raise the sum to $1\%$, or whether the naive
scaling prediction is insufficient.

\subsection{Ties to lighter systems}

Decay or transitions between charmonia provide a lab within which to study
the properties of light mesons. BES~\cite{bes:etainvis}
searched for the decay $\eta(') \to
\mbox{undetectable final states}$, where the $\eta(')$ is produced in
the reaction $J/\psi \to \phi \eta(')$. The $\phi$ as a narrow resonance
is readily identified via its decay into a charged kaon pair, and kinematics
constrain the recoiling $\eta(')$ to a narrow region in the missing momentum.
No signal is seen; an upper limit is placed on the decay of $\eta(')$
to invisible final states relative to decay into two photons, which translate
into absolute branching fractions of $\eta,\ \eta(') \to \mbox{invisible}$ 
of order $10^{-3}$ and $10^{-4}$, respectively.  

\pagebreak
\begin{figwindow}[2,l,%
  \includegraphics*[width=2.5in]{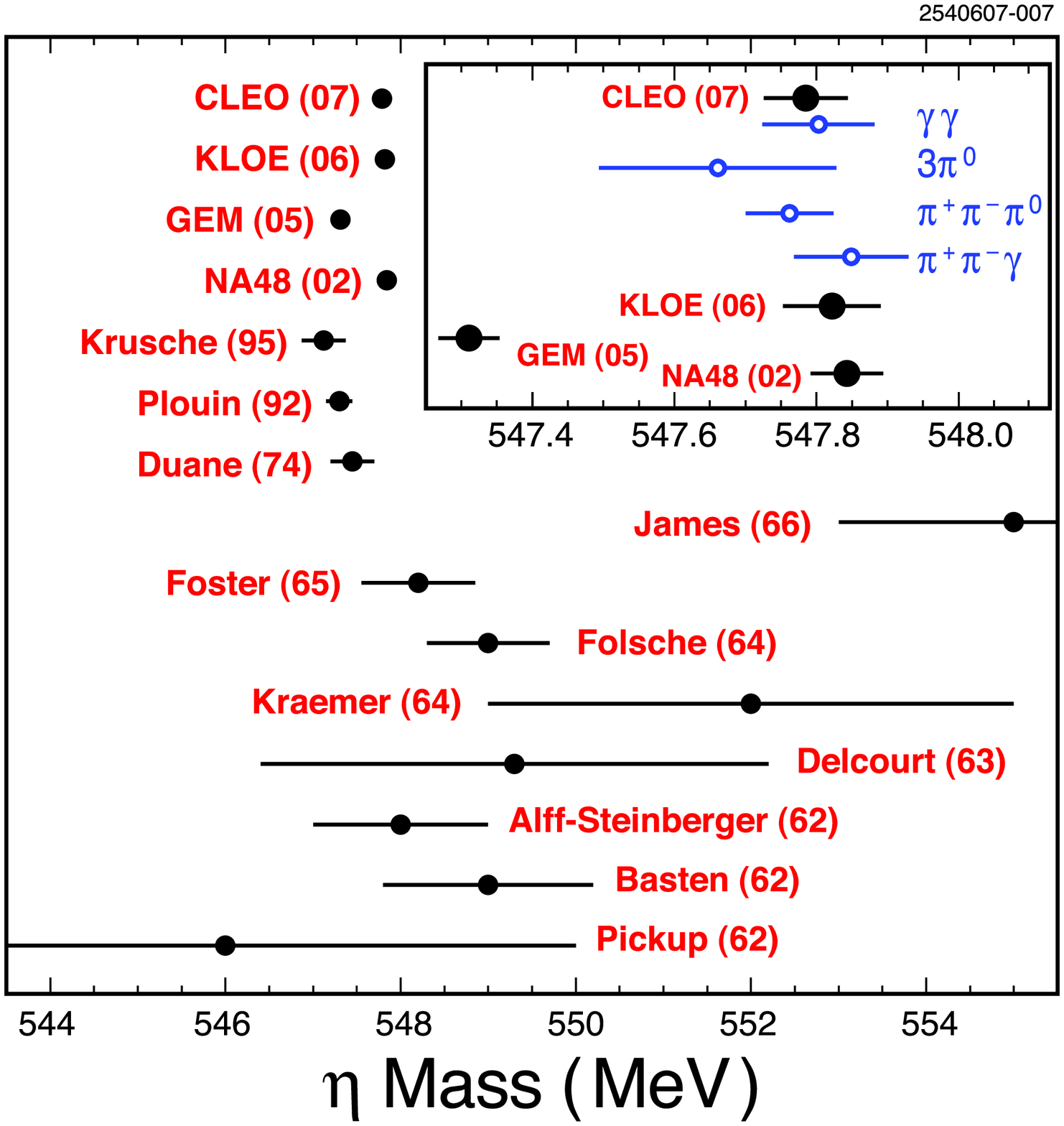},%
  {\label{fig:etamass} World data on the $\eta$ mass.}] 
CLEO used the transition $\psi(2S) \to \eta J/\psi$ with 
$J/\psi \to \ell^+\ell^-$ 
to study the $\eta$ meson.
Branching fractions and ratios thereof~\cite{cleo:etabr} were determined for
$\eta \to \gamma\gamma$, $\pi^+\pi^-\pi^0$, $3\pi^0$, $\pi^+\pi^-\gamma$,
and $e^+e^-\gamma$, a first for such a suite of modes within the same
experiment. Deviations from previous determinations were observed for
$\pi^+\pi^-\gamma$ and $e^+e^-\gamma$ at the level of three standard 
deviations. 
The kinematic properties of the decay and CLEO's resolution
allow to use the invariant mass of the $\eta$ decay products
(except $e^+e^-\gamma$, which has too few events)
to determine the $\eta$ mass~\cite{cleo:etamass} as a by-product.
The precision achieved is
comparable to that of dedicated experiments (Fig.~\ref{fig:etamass}).
\end{figwindow}

\vspace{5mm}

\section{Conclusions}
Charm continues to provide many interesting avenues to enhance the
understanding of the strong interaction, whether as an open-charm
meson or a charmonium state.  
Charm allows to investigate the impact of presence of the strong
force in weak decays.
The system of states provides information about the underlying 
potential, and hence confirming the existence of expected states
and determining spectroscopic parameters is important in both cases. 
Studying transition and decays provides further
insight into the production mechanism and properties of lighter 
hadrons generated in the reaction. Techniques can be refined in
the charm system, where large data samples are available while
external input allows to overconstrain the problems, thereby
affording a calibration ground for theory techniques
which can then be applied to heavier systems. 
It is mandatory to exploit the charm data samples
now for maximal impact at future facilities.

\section{Acknowledgements}

I would like to thank the organizers for their excellent work
in preparing a delightful and interesting conference.
I thank my colleagues for their input and useful
discussion. This work was supported by the National Science
Foundation under contract NSF PHY-0202078.


\begin{thebibliography}{99}
\bibitem{cleo:fd} 
  M.~Artuso {\it et al.}  [CLEO Collaboration],
  %``Improved measurement of B(D+ --> mu+ nu) and the pseudoscalar decay
  %constant f(D+),''
  Phys.\ Rev.\ Lett.\  {\bf 95}, 251801 (2005).
  %[arXiv:hep-ex/0508057].

\bibitem{cleo:fds} 
  M.~Artuso {\it et al.}  [CLEO Collaboration],
  %``Measurement of the decay constant f(D/s+) using D/s+ --> l+ nu,''
  Phys.\ Rev.\ Lett.\  {\bf 99}, 071802 (2007).
  %[arXiv:0704.0629 [hep-ex]].

\bibitem{babar:fds} 
 B.~Aubert {\it et al.}  [BABAR Collaboration],
  %``Measurement of the pseudoscalar decay constant f(D/s) using  charm-tagged
  %events in e+ e- collisions at s**(1/2) = 10.58-GeV,''
  Phys.\ Rev.\ Lett.\  {\bf 98}, 141801 (2007).
  %[arXiv:hep-ex/0607094].

\bibitem{lqcd:fd_fds} 
  E.~Follana {\sl et al.} %, C.~T.~H.~Davies, G.~P.~Lepage and J.~Shigemitsu  
     [HPQCD Collaboration],
  %``High Precision determination of the pi, K, D and D_s decay constants   from
  %lattice QCD,''
  Phys.\ Rev.\ Lett.\  {\bf 100}, 062002 (2008).
  %arXiv:0706.1726 [hep-lat].

\bibitem{cleo:d_kpipienu} 
  M.~Artuso {\it et al.}  [CLEO Collaboration],
  %``Evidence for the Decay D^0 --> K^_pi^+pi^-e^+nu_e,''
  Phys.\ Rev.\ Lett.\  {\bf 99}, 191801 (2007).
  %arXiv:0705.4276 [hep-ex].

\bibitem{lqcd:ff_shape}
  C.~Aubin {\it et al.},
  %``Charmed meson decay constants in three-flavor lattice QCD,''
  Phys.\ Rev.\ Lett.\  {\bf 95}, 122002 (2005).
  %[arXiv:hep-lat/0506030].

\bibitem{cleo:d0_kpi} 
  D.~G.~Cassel,
  %``Hadronic D decays and Dalitz analyses,''
%{\it In the Proceedings of International Conference on Heavy Quarks and Leptons (HQL 06), Munich, Germany, 16-20 Oct 2006, pp 026}
  arXiv:hep-ex/0702021.

\bibitem{babar:d0_kpi} 
  B.~Aubert {\it et al.}  [BABAR Collaboration],
  %``Measurement of the absolute branching fraction of D0 --> K- pi+,''
  Phys.\ Rev.\  Lett. {\bf 100}, 051802 (2007).
  %arXiv:0704.2080 [hep-ex] (subm. to PRL).

\bibitem{ff:mod_pole}
  D.~Becirevic and A.~B.~Kaidalov,
  %``Comment on the heavy --> light form factors,''
  Phys.\ Lett.\  B {\bf 478}, 417 (2000).
  %[arXiv:hep-ph/9904490].  

\bibitem{ff:series}
  R.~J.~Hill,
  %``The modern description of semileptonic meson form factors,''
  %{\it In the Proceedings of 4th Flavor Physics and CP Violation Conference (FPCP 2006), Vancouver, British Columbia, Canada, 9-12 Apr 2006, pp 027}
  arXiv:hep-ph/0606023.

\bibitem{pdg06} 
  W.~M.~Yao {\it et al.}  [Particle Data Group],
  %``Review of particle physics,''
  J.\ Phys.\ G {\bf 33}, 1 (2006) and 2007 partial update for edition 2008.

\bibitem{belle:threebody}
  K.~Abe {\it et al.}  [BELLE Collaboration],
  %``Measurement of the ratio Br(D0 --> pi+ pi- pi0)/Br(D0 --> K- pi+ pi0),''
  arXiv:hep-ex/0610062.

\bibitem{babar:threebody}
  B.~Aubert {\it et al.}  [BABAR Collaboration],
  %``Precise branching ratio measurements of the decays D0 --> pi- pi+ pi0  and
  %D0 --> K- K+ pi0,''
  Phys.\ Rev.\  D {\bf 74}, 091102 (2006).
  %[arXiv:hep-ex/0608009].

\bibitem{cleo:Ds_PP}
  G.~S.~Adams {\it et al.}  [CLEO Collaboration],
  %``Suppressed Decays of D_s^+ Mesons to Two Pseudoscalar Mesons,''
  Phys.\ Rev.\ Lett.\  {\bf 99}, 191805 (2007).
  %arXiv:0708.0139 [hep-ex]. 

\bibitem{focus:d_4pi} 
  J.~M.~Link {\it et al.}  [FOCUS Collaboration],
  %``Study of the D0 --> pi- pi+ pi- pi+ decay,''
  Phys.\ Rev.\  D {\bf 75}, 052003 (2007).
  %[arXiv:hep-ex/0701001].

\bibitem{eichten:opencharm_xsect} 
  E.~Eichten, K.~Gottfried, T.~Kinoshita, K.~D.~Lane and T.~M.~Yan,
  %``Charmonium: Comparison With Experiment,''
  Phys.\ Rev.\  D {\bf 21}, 203 (1980).

\bibitem{belle:opencharm_xsect}
  K.~Abe {\it et al.}  [Belle Collaboration],
  %``Measurement of the near-threshold $e^+e^- \to D^{(*)\pm}{D}{}^{*\mp}$
  %cross section using initial-state radiation,''
  Phys.\ Rev.\ Lett.\  {\bf 98}, 092001 (2007).
  %[arXiv:hep-ex/0608018].

\bibitem{bes:2-5gevscan} 
  M.~Ablikim {\it et al.}  [BES Collaboration],
  %``Determination of the $\psi(3770)$, $\psi(4040)$, $\psi(4160)$ and
  %$\psi(4415)$ resonance parameters,''
  Phys.\ Lett.\  B {\bf 660}, 315 (2008).
  %arXiv:0705.4500 [hep-ex].

\bibitem{bes:psi3770scan}
  M.~Ablikim {\it et al.}  [BES Collaboration],
  %``Measurements of the branching fractions for psi(3770) --> D0 anti-D0,  D+
  %D-, D anti-D and the resonance parameters of psi(3770) and psi(2S),''
  Phys.\ Rev.\ Lett.\  {\bf 97}, 121801 (2006);
  %[arXiv:hep-ex/0605107];
  M.~Ablikim {\it et al.}  [BES Collaboration],
  %``Precison measurements of the mass, the widths of psi(3770) resonance  and
  %the cross section sigma(e+ e- --> psi(3770)) at E(cm) = 3.7724-GeV,''
  Phys.\ Lett.\  B {\bf 652}, 238 (2007).
  %[arXiv:hep-ex/0612056].

\bibitem{e385:psi2Swidth} 
  M.~Andreotti {\it et al.}  [Fermilab E835 Collaboration],
  %``Precision measurements of the total and partial widths of the psi(2S)
  %charmonium meson with a new complementary-scan technique in antip p
  %annihilations,''
  Phys.\ Lett.\  B {\bf 654}, 74 (2007).
  %arXiv:hep-ex/0703012. 

\bibitem{e1_trans}
  E.~Eichten, S.~Godfrey, H.~Mahlke and J.~L.~Rosner,
  %``Quarkonia and their transitions,''
  arXiv:hep-ph/0701208 (accepted by Rev. Mod. Phys.).

\bibitem{bes:radpsi2S}
  M.~Ablikim {\it et al.}  [BES Collaboration],
  %``Measurement of psi(2S) radiative decays,''
  Phys.\ Rev.\ Lett.\  {\bf 99}, 011802 (2007).
  %[arXiv:hep-ex/0612016].

\bibitem{bes:etainvis}
  M.~Ablikim {\it et al.}  [BES Collaboration],
  %``Search for invisible decays of eta and eta' in the processes J/psi -->  Phi
  %eta and Phi eta',''
  Phys.\ Rev.\ Lett.\  {\bf 97}, 202002 (2006).
  %[arXiv:hep-ex/0607006].

\bibitem{cleo:etabr} 
  A.~Lopez {\it et al.}  [CLEO Collaboration],
  %``Measurement of Prominent eta Decay Branching Fractions,''
  % arXiv:0707.1601 [hep-ex] 
  Phys. Rev. Lett. 99, 122001 (2007).

\bibitem{cleo:etamass} 
  D.~H.~Miller {\it et al.}  [CLEO Collaboration],
  %``Measurement of the eta-Meson Mass using psi(2S) --> eta J/psi,''
  % arXiv:0707.1810 [hep-ex] 
  Phys. Rev. Lett. 99, 122002 (2007).

\end{thebibliography}
\end{document}